\documentclass{article}

\begin{document}

\newcommand{\ddx}[1]{\frac{\partial}{\partial x^{#1}}} 
\newcommand{\ddxp}[1]{\frac{\partial}{\partial x'^{#1}}} 
\newcommand{\dydx}[2]{\frac{\partial{#1}}{\partial{#2}}}
\newcommand{\dydxt}[2]{\frac{d{#1}}{d{#2}}}
\newcommand{\DD}{\Delta}
\newcommand{\GG}{\Gamma}
\newcommand{\tp}{\otimes}             
\newcommand{\ww}{\wedge}             
\newcommand{\la}{\langle}  
\newcommand{\ra}{\rangle}  

\newcommand{\vv}[1]{{\bf #1}}             
\newcommand{\tpp}{\tp\cdots\tp}       
\newcommand{\LLL}{{\mathcal L}}
\newcommand{\f}{{\mathcal F}}
\newcommand{\G}{{\mathcal G}}
\newcommand{\W}{{\mathcal W}}
\newcommand{\M}{{\mathcal M}}
\newcommand{\LL}{\Lambda}
\newcommand{\ppp}{\partial}
\newcommand{\til}[1]{\stackrel{\sim}{#1}} 

\newcommand{\be}{\begin{eqnarray}}
\newcommand{\ee}{\end{eqnarray}}
\newcommand{\bes}{\begin{eqnarray*}}
\newcommand{\ees}{\end{eqnarray*}}

\title{Poincare-Cartan form for scalar fields in curved background.}
\author{Pankaj Sharan
\\Physics Department, Jamia Millia Islamia,\\ New Delhi 110 025, INDIA}
\date{}

\maketitle

\begin{abstract} 
Poincare-Cartan form for scalar field is constructed as a differential 4-form in a `directly Hamiltonian' formalism which does not use a Lagrangian. The canonical momentum $p$ of a scalar field $\phi$ is a 1-form and the Poincare-Cartan 4-form $\Theta$ is $(*p)\wedge d\phi-H$ where the Hamiltonian $H$ is a suitable 4-form made from $\phi$ and $p$ using the Hodge star operator defined by the Riemannian metric of the background spacetime. An allowed field configuration is a 4-dimensional surface in the 9-dimensional extended phase space such that its tangent vectors annihilate $\Omega=-d\Theta$. Relation of this to variational principle, symmetry fields and conserved quantities is worked out. Observables are defined as differential 4-forms constructed from field and momenta smeared with appropriate test functions. A bracket defined by Peierls long ago is found to be the suitable candidate for quantization.
\end{abstract}

\section{Introduction}
The Hamiltonian formulation is basic to quantum theory. Despite this, the quantum
theory of {\em fields} always begins with a Lagrangian. The main reason for this is,
in words of P. A. M. Dirac \cite{dirac}
``it is not at all easy to formulate the conditions for a theory to be 
relativistic in terms of the Hamiltonian''.  The conditions for relativistic
invariance are satisfied by choosing a Lagrangian function to be a relativistic scalar.
This function can be constructed as a scalar by
balancing indices of vector, tensor or spinor fields and their four-dimensional derivatives.

The purpose of this series of papers is to show how one can directly set up a Hamiltonian
formalism for relativistic fields, including fields in arbitrary curved background,
without first writing a Lagrangian and then proceeding to the Hamiltonian 
through the Legendre transformation. A Hamiltonian formalism can be set up in terms of fields and their
canonical momenta quite as easily as a Lagrangian is written in terms of fields and
their derivatives provided we treat fields and canonical momenta as differential forms (with values
in spaces that characterize them). The canonical momenta in our formalism are differential
forms of {\em one degree higher than the fields}. Thus, mathematically,  coordinates and their momenta 
are not quantities of the same type. This fundamental change in the mindset
allows us to set up a covariant coordinate-free formalism which is Hamiltonian
from the very beginning and does not require a Lagrangian for its definition.
 
Preliminary work in this direction already exists in formalisms variously 
known as `finite-dimensional covariant
formalism' or `multisymplectic' or `polysymplectic' formalism. The basic idea
was given by Weyl and de Donder in the so-called `multiple integral problem in the calculus of variations'    
and was developed by Kastrup, Kanatchikov, Gotay et al and Rovelli and others. See references and
discussion at the end of this section.

It is commonly believed that the Hamiltonian formalism singles out time as a 
special variable and this spoils the relativistic invariance which would have
required space and time to be treated
on the same footing. This is true if one regards derivative of fields
with respect to the time coordinate as fundamentally different from that with respect
to a space coordinate.
But if we treat all the four derivatives $\ppp_\mu\phi$ of 
a scalar field $\phi$  as one quantity then it follows we
should allow four components $p_\mu$ of momenta to be associated with this
one field variable.
We should not pair one coordinate with one momentum 
degree of freedom. Such a pairing 
is a peculiarity of the Hamiltonian mechanics based on 
a single evolution parameter time,
whereas fields extend in space as well as time. 

The fundamental principle in classical mechanics is that variation of a quantity
called action is zero. The laws of nature allow only those configurations of physical
variables which achieve an extremum for action. And this requirement of extremum is
the classical limit when $\hbar$ is regarded as small. 

For classical mechanics  time can be regarded as
a `base manifold' and coordinates and momenta are in the `fibre'. This is 
the extended phase space. Action is an integrated 
value $\int \Theta$ of a one-form called Poincare-Cartan form $\Theta=pdq-Hdt$
on a supposed trajectory in extended phase 
space. The variation of the trajectory is determined by a vector field of infinitesimal displacements.
The condition that the Lie derivative of the action along the proposed
trajectory with respect to the field of
variation is zero when the the coordinates are fixed at the ends of the trajectory
determines the trajectory. The tangent vectors to allowed trajectories determine  
Hamiltonian vector fields. 
The variational principle can be reformulated by saying that the 
Hamiltonian vector fields of allowed trajectories annihilate 
the differential 2-form $\Omega=-d\Theta$ where $\Theta$ is the 
Poincare-Cartan (PC) form.\cite{bundle}

In order to set up a purely Hamiltonian formalism for fields, we must 
first try to define a suitable PC-form for fields.

The PC-form $\Theta$ has two parts : the so-called fundamental 
form $pdq$
which governs geometry of the phase space, and the Hamiltonian part
$-Hdt$ which determines the dynamics for the given system.   

The logic for writing a PC-like form for a scalar field goes like this. 
In field theory, the field $\phi$ is the configuration 
variable analogous to $q$. Time and space are four ``time'' variables $t^\mu,\mu=0,1,2,3$. 
We purposely use the 
letter $t$ also for space coordinates $t^i,i=1,2,3$ to emphasize this point. 
We expect the PC-form for fields to be a differential
{\em four}-form whose spacetime integral will give the quantity we call action. 
For a single scalar field, the momenta are related to velocities 
by $p_\mu=\ppp_\mu\phi$. Thus we can keep them together as a 
1-form $p=p_\mu dx^\mu$. To 
get a fundamental 4-form similar in appearance to $pdq$ we need a 
3-form (and not a 1-form $p$) to be multiplied to $d\phi$. 

There is a natural way to produce a 3-form out of a 1-form, 
namely, by using the metric of
the spacetime through the star-dual $*p$. We are led naturally 
to introduce the 
following expressions for the PC 4-forms :
\be \Theta=(*p)\ww d\phi-H \ee
where $H$ is a differential 4-form
\be H &=& \frac{1}{2}(*p)\ww p+\frac{1}{2}m^2\phi^2(*1)\nonumber\\
&=& \left(-\frac{1}{2}\la p,p\ra+\frac{1}{2}m^2\phi^2\right)(*1).
\ee
We have used the definition of the star operator relating it to
the inner product determined by $g_{\mu\nu}$ which has a 
signature corresponding
to $(-,+,+,+)$. Our convention for the star operator 
is the same as Sharan\cite{sharan} or
Choquet-Bruhat and DeWitt-Morette\cite{AMP}, and is very briefly summarized in Appendix A.

Observe that the Hamiltonian 4-form $H$ is defined solely in terms of the 
field variable $\phi$ and the momenta $p$ (or $*p$).
It is a coordinate independent definition. $H$ is a 4-form and it should
not be confused with the
Hamiltonian density or the energy density of the 
usual Lagrangian field theory. (That
density is a 3-form which will be seen to be the conserved 
quantity for time translations in static spacetimes.)

A field theory involves infinitely many degrees of
freedom. The traditional view is to think of each value $\phi(\vv{x},t)$ for space points $\vv{x}$
on a plane of constant time $t$ as a separate degree of freedom for a scalar field. This is 
the usual `3+1' Hamiltonian point of view. 
See Chernoff and Marsden\cite{chernoff} for a rigorous account of Hamiltonian
systems of infinitely many degrees of freedom. 

There is another, more interesting way to look at this.
One can regard a solution
of the field equations as a section or a surface in the 
{\em finite} dimensional 
extended phase space four of whose coordinates are the spacetime
coordinates. The infinitely many ways
in which this surface can be embedded in the extended phase 
space is a reflection of the infinitely many degrees of freedom of the field system.

For our example, the extended phase space for a single scalar field 
is a nine-dimensional manifold (four spacetime variables $t^\mu$, one 
field variable $\phi$ and four momentum variables in $p$). A possible 
configuration of the field (that is, a solution of the field equations)
is a four dimensional surface in this nine dimensional space
``above'' the four dimensional spacetime. The fiber bundle picture is helpful
because we are interested in `sections' or functions from
spacetime base into the fields and momenta. Mathematically, there may
be more general submanifolds or surfaces in the extended phase space but they do not
seem to be physically relevant.

As mentioned above the mathematical formalism of the present paper is 
similar to the ``multisymplectic'' Lagrangian approach to field theory in the works 
of Le Page, as reviewed and developed by Kastrup\cite{kastrup}, the De Donder-Weyl\cite{rund}
approach of Kanatchikov\cite{kanat} and the covariant
Hamiltonian-Jacobi formalism of Rovelli\cite{rovelli}. Recent contributions to multisymplectic formalism
are by Gotay and collaborators\cite{gotay}.
Our approach is different from these because
we use the background spacetime metric
in an essential way through the Hodge star operator. Also, we treat the
spacetime degrees of freedom $t^\mu$ which specify the base differently 
from the field or momentum degrees
of freedom which are in the fibre above the base.
We require the PC-form to be a 4-form whose first term is linear in $d\phi$ to imitate $pdq$ term
and the second term is a 4-form $-H$ proportional to volume form $(*1)$. 
If, for instance, there are two fields $\phi_1$ and $\phi_2$, a 
4-form involving a factor $d\phi_1\ww d\phi_2$ is possible in principle but that
does not seem be allowed in the formalism for matter fields. Similarly 
other `non-canonical' expressions are possible in place of the standard $pdq-Hdt$ like
expression. For gravity, the Einstein-Hilbert
PC form does seem to have a non-standard expression as we shall see in a later paper.
But gravity is a special case anyway. 
For gravity the `internal' degrees of freedom in the fibre related to arbitrary choice of
local inertial frames and spacetime bases which define the 
transformation of all field and momenta differential forms happen to coincide.  

It is natural and tempting to put our formalism in the fibre bundle language, but we avoid
that for the sake of clarifying the physical concepts. For most part we assume the 
bundle to be a direct product of spacetime and the fibre manifold.

Our aim is to develop a purely Hamiltonian approach and define 
a suitable bracket to help build a quantum theory. The only reliable way to
convert a classical theory into a quantum theory is to define a suitable antisymmetric 
(or symmetric) bracket for observables of the theory which can be re-interpreted 
in quantum theory as a commutator
(or anticommutator). Our phase space has a very different character
than the traditional phase space and our coordinate and momenta are differential
forms of different degrees. In the traditional formalism the 
observables are real valued functions on the phase space
and the definition of the Poisson bracket uses the pairing of one coordinate 
with one canonical momentum degree of freedom. But that is special to one-time
formalism of mechanics.

But in mechanics there is another way to look at the Poisson bracket. The bracket $\{B,A\}$
of two observables $A$ and $B$ refers to the rate of change of one observable $B$
when the other observable $A$ acts as the Hamiltonian. In one-time formalism
the rate of change of a quantity is mathematically the same type of quantity as the original quantity.
When space {\em and} time are evolution parameters then the rate of change can only mean
rate of change  along a vector field. This rate of change is
the Lie derivative. Thus we need the Lie derivative of one quantity 
with respect to the Hamiltonian vector field 
determined on the phase space by the other quantity.

Whereas the Hamiltonian vector field for any observable
exists for in mechanics the same may not be so for fields. We find that the
concept of a covariant bracket introduced by Peierls\cite{peierls} in 1952 (and promoted extensively by
De Witt\cite{dewitt}) is a natural  object to use in our Hamiltonian theory of fields. Here
the rate of change of one quantity is taken when {\em the other quantity is added to the 
Hamiltonian as an infinitesimal perturbation} and vice-versa. The Poisson brackets of mechanics can be 
defined without reference to any Hamiltonian whereas the Peierls bracket
requires the existence of a suitable governing Hamiltonian. Roughly speaking, the Poisson bracket 
can be described as the ``equal time'' Peierls bracket with zero Hamiltonian. 

This gives us added insight into the Hamiltonian mechanics of one time formalism,
particularly the concept of causality in systems with time dependent Hamiltonians.
The interesting features for one-time formalism of classical mechanics 
relating to causality and Peierls bracket which  are revealed by our formalism of fields
will be published elsewhere.

In section II we define the Poincare-Cartan form. We set up the variational 
principle and Noether's theorem in sections III and IV. We define our 
observables as smeared 4-forms and their 
Peierls bracket in section V. Symmetries and conserved quantities are 
discussed in section VI and VII and the Hamilton-Jacobi
formalism is discussed briefly in section VIII. Notation is summarized in appendix A.
A calculation for the solution manifold in section II is outlined in appendix B. 

\section{Poincare-Cartan form for a scalar field}

For fields the extended phase space is a bundle with the four-dimensional spacetime $T$ as base space.
We denote the spacetime points by $t=(t^0,t^1,t^2,t^3)\in T$. Let us consider the one-dimensional
fibre of 0-forms with coordinate $\phi$ and the four-dimensional fibre of 1-forms 
whose points are labelled by $p=p_\mu dt^\mu$. We can think of the extended phase space $\GG$ to
be the base (of spacetime) with a five-dimensional fibre at each point which is a direct 
sum of 0-forms and 1-forms. 

We require the momentum canonical to a scalar field  $\phi$ to be a 1-form $p=p_\mu dt^\mu$
where coefficients $p_\mu$ are independent variables.
The PC-form on this nine-dimensional extended phase space 
(with coordinates $t^\mu,\phi,p_\mu$) is chosen as
\be \Theta=(*p)\ww d\phi- H \ee
where $H$ is a 4-form constructed from $p$ and $\phi$. The simplest
choice is a Hamiltonian with a `kinetic energy term' and a `mass term' :
\be H &=& \frac{1}{2}(*p)\ww p+\frac{1}{2}m^2\phi^2(*1)\nonumber\\
&=& \left(-\frac{1}{2}\la p,p\ra+\frac{1}{2}m^2\phi^2\right)(*1).
\ee
It is necessary to point out here that although our star operator 
is limited to the four-dimensional spacetime the exterior derivative works in
the nine-dimensional extended phase space. 
Thus $d\phi$ is linearly independent of
$dt^\mu$ and so also independent of $p=p_\mu dt^\mu$. 
The coefficients $p_\mu$ are independent coordinates. Therefore
$dp_\mu$ are linearly independent of $d\phi$ and $dt^\mu$.

It is also worth pointing out that the definition of star operator requires
the existence of a set of orthonormal basis fields with a given orientation.
this is where gravity sneaks in as a universal field. In the present paper
the gravitational field will be fixed as an external field defining the star operator. 

Dynamics is determined by the 5-form
\be \Omega = -d\Theta= -(d*p)\ww d\phi + dH, \ee
and the variational principle can be stated as follows :
\begin{quote}
The solution manifold $\sigma$ in the extended phase space 
is a section whose tangent vectors annihilate $\Omega$.
\end{quote}
This statement is explained below. The relation of this  statement of variational principle to
the usual statement for variation of the action is discussed in the next section.

In mechanics we look for phase trajectories. Here, in field theory we look for a four-dimensional
image of a {\em section}, that is, a mapping $\sigma$ from the four dimensional 
base into a 4-dimensional submanifold of the nine-dimensional extended phase space :
\be \sigma : t=\{t^\mu\} \to \{t^\mu,\phi=F(t), p=G_\mu(t)dt^\mu\}. \ee
By abuse of language we will denote the mapping as well as its image of the base
by the same letter $\sigma$. The context will make it clear what the symbol
corresponds to.

$\sigma$ defines a surface or sub-manifold such
that if $X_0,X_1,X_2,X_3$ are four linearly independent vectors in the 
tangent space of this submanifold at any point then the 1-form
obtained by the interior product of all these with $\Omega$ should be zero :
\be i(X_3)i(X_2)i(X_1)i(X_0)\Omega = 0.\ee
Recall that the interior product of a vector field $X$ with an $r$-form $\alpha$
is defined as the $(r-1)$-form $i(X)\alpha$ so that 
$i(X)\alpha(Y_1,...,Y_{r-1})=\alpha(X,Y_1,...,Y_{r-1})$.
Depending on typographical convenience we shall denote the 
interior product of a vector field $X$ with a form $\alpha$ by
$i(X)\alpha$ or $i_X\alpha$.

The meaning of variational principle above is that for arbitrary vector field $Y$ on $\GG$, 
\be \Omega(X_3,X_2,X_1,X_0,Y)=0.\ee
In the following we call $\sigma$ determined by this condition as a ``solution submanifold''.
We can choose $X_\mu$ to be just the push-forwards by $\sigma$
of the coordinate basis vectors $\ppp_\mu\equiv \ppp/\ppp t^\mu$ :
\bes X_\mu=\sigma_*(\ppp_\mu)=\ppp_\mu+F_{,\mu}\ppp_\phi+G_{\nu,\mu}\ppp_{p_\nu}. \ees

For our case $\Omega$ can be calculated
easily. Using
\bes d(*p\ww p)=d(*p)\ww p+*p\ww(dp)=2d(*p)\ww p \ees
we get
\bes dH=(d*p)\ww p+m^2\phi\, d\phi\ww(*1) \ees
Substituting in $\Omega$ we see that it {\em factorizes}
\be \Omega = (d*p-m^2\phi(*1))\ww(p-d\phi), \ee
where we use the fact that the 5-form $(*1)\ww p$ 
in four variables $t$ is zero because there are five factors of $dt$'s.

We give details of the calculation for flat space in Appendix B. 
The condition on $F$ and $G_\mu$ to define  a solution manifold is 
\be G_\mu=F_{,\mu},\qquad d*dF-m^2F(*1)=0 \ee
which is the solution $\phi=F(t)$ to the Klein-Gordon equation for the field $\phi$.  

There is a less rigorous but physically straightforward way to see what solution manifold
should be. Vector fields annihilating
$p-d\phi$ imply $p_\mu=\ppp_\mu\phi$. Similarly, for $p=d\phi$,
the first factor gives zero if
\be d*d\phi-m^2\phi(*1)=0 \ee
Now 
\be d*d\phi &=& \ppp_\mu(\sqrt{|g|}g^{\mu\nu}\ppp_\nu\phi)
dt^0\ww\dots\ww dt^3\nonumber\\
&=& \frac{1}{\sqrt{|g|}}\ppp_\mu(\sqrt{|g|}g^{\mu\nu}\ppp_\nu\phi)(*1)
\ee
and thus $\phi$ satisfies the Klein-Gordon equation 
with the Laplace-Beltrami operator
of the curved space.

We close this section with a few remarks. 
\begin{enumerate}
\item
Since $\phi$ and $t^\mu$ are coordinates in the extended
phase space, $d\phi$ and $dt^\mu$ are linearly independent.
Therefore $p_\mu dt^\mu-d\phi=0$ is meaningless as it stands. 
What it implies is that 
there exists a subspace or submanifold $\sigma$ of the nine-dimensional 
extended phase space
such that any of the independent vector fields $X_\mu$ tangent to $\sigma$ satisfies 
\bes (p_\mu dt^\mu-d\phi)(X)=0.\ees 
The 1-form $p_\mu dt^\mu$ has non-zero coefficients for the $dt^\mu$'s
and zero for $d\phi$ and $dp_\mu$.
These coefficients $p_\mu$ themselves  are independent coordinates.
Thus, although $*p\ww p=-\la p,p\ra (*1)$ is proportional to
4-form $*1$ its exterior derivative $d(*p\ww p)$ need not be zero.\\

\item
If there are several fields $\phi^a$ then we can construct the PC-form similarly
as
\be \Theta = *p_a\ww d\phi^a-H \ee
where $p_a=p_{a\mu}dt^\mu$ are canonical momenta for the fields $\phi^a$ and
$H$ is a 4-form depending on all the fields and the momenta. \\

\item
If $(\phi_1,p_1)$ and $(\phi_2,p_2)$ are two solutions for the scalar field,
then the 4-form
\bes &&d(\phi_1 *p_2-\phi_2 *p_1) \\
&=& d\phi_1\ww *p_2+\phi_1 d*p_2-(1\leftrightarrow 2)\\
&=& (d\phi_1)\ww(*d\phi_2) +\phi_1(m^2\phi_2)*(1)-(1\leftrightarrow 2)\\
&=& (d\phi_1)\ww(*d\phi_2)-(d\phi_2)\ww(*d\phi_1)\\
&=& 0
\ees
because, (using the identity $(*t)\ww s=(-1)^{r(n-r)}t\ww(*s)$
for any $r$-forms $t$ and $s$ in an $n$-dimensional space) we conclude that
in our case
\bes (d\phi_1)\ww(*d\phi_2)=-(*d\phi_1)\ww(d\phi_2)=(d\phi_2)\ww(*\phi_1).\ees
Thus, by Stokes theorem the integral
\bes \oint (\phi_1 *p_2-\phi_2 *p_1) \ees
over any closed surface is zero. This leads to a linear space of solutions
on which there is a time-independent scalar product.
\end{enumerate}

\section{Stationary Action}

We have seen that a specific solution to the field equations can be realized as 
a four-dimensional submanifold $\sigma$ of the nine-dimensional extended phase space.

Hamilton's variational principle involves comparing the integral of
the PC-form on a proposed four-dimensional solution submanifold
with a similar integral on a neighboring submanifold.

Let $\sigma : t\to \{ \phi=F(t),p_\mu=F_{,\mu}\}\in \GG$ be the 
submanifold corresponding to some given solution.

Let $D$ be a region of spacetime and $\ppp D$ its boundary. Calculate the PC-form
$\Theta$ on the region $\sigma(D)$ of the extended phase-space mapped by $\sigma$. 

Let $Y$ be a vector field of variation. We can paraphrase Arnold's elegant argument\cite{arnold}
for mechanics and apply to fields. Calculate the Lie derivative using the formula
$L_Y=i_Y\circ d+d\circ i_Y$ (see for example \cite{michor}) :
\bes \delta_Y \int_{\sigma(D)}\Theta & \equiv & L_Y\int_{\sigma(D)}\Theta\\
&=& \int_{\sigma(D)}L_Y\Theta\\
&=& \int_{\sigma(D)}(i_Y\circ d+d\circ i_Y)\Theta\\
&=& -\int_{\sigma(D)}i_Y\Omega +\int_{\sigma(D)}d[i_Y\Theta]\\
&=& \oint_{\ppp \sigma(D)}i_Y\Theta
\ees
where the integral of $i_Y\Omega$ on the submanifold $\sigma$ is zero 
because the integral evaluates $i_Y\Omega$ on tangent vectors $\sigma_*(\ppp_\mu)$
to the proposed solution sumanifold 
which is zero. Thus variational principle can also be expressed as, 
\be \delta_Y \int_{\sigma(D)}\Theta=\left.\oint_{\ppp \sigma(D)}i_Y\Theta\right|_{0}.\ee
Here we use the symbol $0$ to denote a quantity ``on-shell'', that is,
evaluated on a solution submanifold.
It needs to be emphasized that since the variation field $Y$ is not restricted to
the solution surface, it will be a mistake to use $\phi=F,p_\mu=F_{,\mu}$
{\em before} the evaluation of $i_Y\Theta$. 

Since $\Theta$ involves $d\phi$ and $dt^\mu$ (and no $dp_\mu$),
the surface integral of 3-form $i(Y)\Theta$ gives zero if the infinitesimal field
$Y$ is zero along the directions $\ppp/\ppp t^\mu$ and $\ppp/\ppp \phi$. But there is no
restriction on variation in momenta directions.

We can re-express the variational principle (or principle of stationary action)
in extended phase space as :
\begin{quote}
\textit{ 
Under variation by a field $Y$ 
with $\phi,t^\mu$ held fixed at the boundary 
the action 
evaluated at the solution submanifold $\sigma$ 
is stationary :
}
\be \delta_Y \int_{\sigma(D)}\Theta=0.\ee
\end{quote}

\section{Noether's Theorem}

Let us consider a variation $Y$ not necessarily zero at the boundary $\sigma(D)$
where $\sigma$ is solution manifold.
Equation (14) for variations is 
\be \delta_Y \int_{\sigma(D)}\Theta=\int_{\sigma(D)}  L_Y\Theta
=\left.\oint_{\ppp \sigma(D)}i_Y\Theta\right|_{0} \ee

If we know that for some given type of variation $Y$, 
\be  L_Y\Theta =0 \ee
then we say that action in invariant under the infinitesimal
mapping represented by the fields $Y$ and $Y$ is called a `symmetry field'. 
Usually, the symmetry fields satisfy the conditions $L_Y(*p\ww d\phi)=0$ and 
$L_Y H=0$ separately.
The surface integral 
\be \left.\oint_{\ppp \sigma(D)}i_Y\Theta\right|_{0}=0 \ee
gives a conservation law for the 3-form $i_Y\Theta$. 
In the particular case when the boundary $\ppp D$ is
constituted by two spacelike surfaces, the 3-form $i_Y\Theta$, restricted to either 
surface represents the volume density of the conserved ``charge'' on that surface.

\section{Observables and Peierls bracket}

Our formalism treats coordinate $\phi$ and its canonical momentum $p$ respectively
as 0- and 1-forms. In classical mechanics they seem to be quantities of the 
same type because in one-dimensional base manifold representing time, 
0-forms and 1-forms are both 1-dimensional spaces. 
This situation changes for field theory in four dimensions. There 0- and 1-forms 
are respectively spaces  of one and four dimensions.

The observables of our theory are quantities like action : integrated quantities  
over a four dimensional submanifold. A typical observable is an 
integrated 4-form $A=\int \alpha$.
The support of $\alpha$, that is set over which it has non-zero values could be  
suitably restricted to allow for local quantities as observables. For
example, the scalar field $\phi$ is related to the observable
$\int \phi j (*1)$   where $j(t)$ is a scalar `switching function' 
which is non-zero in a small spacetime region. 
For simplicity we would call both the
integrated as well as the non-integrated quantity by the same name `observable',
and it leads to no confusion.

The Peierls bracket is the natural bracket-like quantity in 
this formalism. When the Hamiltonian 4-form $H$ is perturbed by observable $\lambda B$
(where $\lambda$ is an infinitesimal parameter)
the solution manifold shifts, and, after taking causality into account,
the difference between the two solutions at different points in the limit of $\lambda\to 0$ 
determines a `vertical' vector field $X_B$. This field
changes all other observables. The change in an observable
$A$ is equal to the Lie derivative $D_BA\equiv L_{X_B}A$ of $A$ with respect to $X_B$. 
Switching the roles of $B$ and $A$ we can calculate $D_AB$. The
Peierls bracket $[A,B]$ is defined as the difference $D_BA-D_AB$. 

For illustration we outline the calculate the Peierls bracket for the scalar field
with itself in Minkowski space.
The observable in question is the integrated 4-form 
\bes B=\int \beta=\int \phi j (*1) \ees 
where $j$ is a switching function in spacetime with which the field $\phi$ is `smeared'. The 
Hamiltonian is changed to $H+\lambda B$ and the solution
manifold given by $t\to \phi=F_0(t),p_\nu=F_{0,\nu}$ gets
modified to a solution manifold which is determined by the 5-form
\bes \Omega_B &=& -d(*p)\ww d\phi +dH +\lambda d\phi j (*1)\\
&=& [d(*p)-m^2\phi(*1)-\lambda j(*1)]\ww [p-d\phi]. \ees
No derivative of $j$ appears because that would involve five factors of $dt$'s
and there can be only four such factors in a wedge product.
The equations for a solution $t\to \phi=F(t),p_\nu=G_\nu$ become
\bes G_\nu=F_{,\nu},\qquad (\ppp^\mu\ppp_\mu -m^2)F = \lambda j. \ees
The modification caused by $\lambda B$ as $\lambda \to 0$ to the solution 
$F_0$ is given by the retarded solution to the inhomogeneous Klein-Gordon equation,
\bes F(t)=F_0(t)+\lambda K(t),\qquad G_\nu=F_{,\nu} \ees 
where
\bes K(t)=\int G_R(t-s)j(s)d^4s. \ees
The retarded and advanced Green's functions $G_R(t),G_A(t)$ are the unique solutions
\bes G_{R,A}(t)=\frac{1}{(2\pi)^4}\int d^4k\, 
\frac{\exp(-ik^0t^0+i\vv{k}\cdot\vv{t})}{(k^0\pm i\epsilon)^2-\vv{k}^2-m^2}\ees
of
\bes (\ppp^\mu\ppp_\mu -m^2)G_R(t) =\delta^4(t)\ees
with the boundary condition that $G_R(t)$ is non-zero only in the forward light-cone
and $G_A(t)$ in the backward light-cone.

Thus the vertical field is determined to be ($\lambda\to 0$ can be factored out
to give the tangent vector field)
\bes Y_B=K(t)\dydx{}{\phi}+K_{,\nu}\dydx{}{p_\nu} \ees

Consider the observable 
\bes A =\int \alpha=\int \phi k (*1) \ees 
where $k(t)$ is another switching function. The change in $A$
due to $B$ is given by $D_BA =L_{Y_B}(A)$. Now,
\bes L_{Y_B}(A) &=& \int [i_Y (d\phi k (*1))+d(\phi k \,i(Y)(*1))]\\
&=& \int k K (*1), \ees
because $i(Y_B)(*1)=0$. Thus 
\bes D_BA &=&\int d^4t k(t)K(t)\\
&=& \int\int d^4t d^4s k(t)G_R(t-s)j(s) 
\ees
Reversing the role of $B$ and $A$ we get the Peierls bracket
\bes [A,B]=D_BA-D_AB=\int\int d^4t d^4s k(t)\Delta(t-s)j(s) \ees
where $\Delta$ is the Pauli-Jordan function $\Delta=G_R-G_A$.
This is equivalent to the commutator
\bes [\phi(t),\phi(s)]=\Delta(t-s) \ees
when $k$ and $j$ are Dirac deltas with support at $t$ and $s$ respectively. 

The Peierls bracket for the field $\phi$ and momentum $p$ can be 
calculated by considering the observable 
\bes C=\lambda (*p)\ww l=-\lambda p_\mu l^\mu (*1)\ees
where in this case we
must employ a 1-form switching function $l$ to smear the momentum. The
5-form is 
\bes \Omega_C= [d(*p)-m^2\phi(*1)]\ww[p+\lambda l-d\phi]. \ees
The relevant equation for the modified solution is 
\bes (\ppp^\mu\ppp_\mu -m^2)F = \lambda \ppp^\mu l_\mu \ees
because $d(*p)$ becomes $d(*(d\phi-l))=\ppp^\mu\ppp_\mu\phi-\ppp^\mu l_\mu$.
The change in $B$ is
\bes D_CB &=& \int\int d^4t d^4s j(t)G_R(t-s)(\ppp^\mu l_\mu)(s). \ees
On the other hand we have already calculated the vertical field for $B$
which gives
\bes D_BC &=&-\int K_{,\mu}l^\mu (*1)\\ 
&=& -\int\int d^4t d^4s\, l^\mu(t) \ppp_{t^\mu} G_R(t-s)j(s)\\
&=& \int\int d^4t d^4s (\ppp_\mu l^\mu)(t)  G_R(t-s)j(s)\\
&=& \int\int d^4t d^4s  j(t)G_A(t-s)(\ppp_\mu l^\mu)(s)\ees
after integrating by parts in the third step.
Therefore, 
\bes [B,C]=\int\int d^4t d^4s  j(t)\Delta(t-s)(\ppp_\mu l^\mu)(s)\ees
which, for $j(t)=\delta^4(t)$ and $l_\mu=(1,0,0,0)\delta^4(s)$ gives the 
equal-time ($t^0-s^0$) canonical Poisson bracket
of the ``3+1'' version of field theory
\bes [\phi(t,\vv{t}),p_0(t,\vv{s})]=\delta(\vv{t}-\vv{s})\ees
because
\bes \ppp_0\Delta(t)=-\delta^3(\vv{t}). \ees

\section{$i_Y\Theta$ for $Y=v^\mu\ppp/\ppp t^\mu$}
As an illustration of the Noether theorem in our formalism
let us evaluate $i_Y\Theta$ for the present scalar field case 
for spacetime translations. The vector field for constant infinitesimal
displacement $v^\mu$ is
\bes Y=v^\mu\dydx{}{t^\mu}\ees
We are not assuming that spacetime is flat or that $Y$ are Killing fields of translation
symmetry.

We know that 
\bes *p &=& p_\mu *(dt^\mu)\\
&=& \frac{1}{3!}\sqrt{-g}p_\mu g^{\mu\alpha}\varepsilon_{\alpha\nu\sigma\tau}
(\nu\sigma\tau)\\
&\equiv & \frac{1}{3!}\sqrt{-g}p^\alpha\varepsilon_{\alpha\nu\sigma\tau}
(\nu\sigma\tau)
\ees
where we introduce a convenient notation
\bes (\nu\sigma\tau) \equiv dt^\nu\ww dt^\sigma\ww dt^\tau, \ees
with similar notation for two or four factors of $dt^\mu$ and we have defined the
contravariant canonical momentum 
\bes p^\mu=g^{\mu\nu}p_\nu .\ees

A simple calculation using 
\bes i_Y(dt^\mu\ww dt^\nu\ww dt^\sigma) &=& v^\mu(dt^\nu\ww dt^\sigma)
-v^\nu(dt^\mu\ww dt^\sigma)\\&&+
v^\sigma(dt^\mu\ww dt^\nu)\ees
gives,
\bes i_Y(*p)=\frac{1}{2!}\sqrt{-g}p^\alpha v^\beta
\varepsilon_{\alpha\beta\sigma\tau}(\sigma\tau).\ees
We can write this  also as 
\bes i_Y(*p)= p_\mu v_\nu *(dt^\mu\ww dt^\nu)= *(p\ww Y^\flat)\ees
where $v_\mu=g_{\mu\nu}v^\nu$ and $Y^\flat=v_\nu dt^\nu$ is the covariant field
corresponding to $Y$ after lowering the index by the metric.

As $Y$ involves $\ppp/\ppp t^\mu$ whose action on $d\phi$ is zero
\bes i_Y(*p\ww d\phi)=(i_Y*p)\ww d\phi, \ees
and 
\bes i_Y[*p\ww p\,]&=&[i_Y*p]\ww p-*p(i_Yp)\\
&=&*(p\ww Y^\flat)\ww p-p(Y)*p.
\ees

The formula $i_Y(*p)=*(p\ww Y^\flat)$,
although elegant, is not very useful for calculations. A straightforward expression 
for $i_Y(*p)\ww p$ is
\bes i_Y(*p)\ww p=[p_\mu(p.v)-v_\mu(p.p)]*(dt^\mu) \ees
where
\bes p.v=p_\mu v^\mu=\la p,Y^\flat\ra,\qquad p.p= p_\mu p^\mu=\la p,p\ra. \ees

Thus the calculation of $i_Y\Theta$ proceeds as follows,
\bes i_Y\Theta&=&i_Y\left[*p\ww d\phi-\frac{1}{2}*p\ww p
-\frac{1}{2}m^2\phi^2*(1)\right]\\
&=&(i_Y*p)\ww\left(d\phi-\frac{1}{2}p\right)+\frac{1}{2}p(Y)*p\\
&&-\frac{1}{2}m^2\phi^2i_Y*(1)
\ees
Evaluating it ``on-shell'' means we can put $p=d\phi$. Using expression for 
$i_Y(*p)\ww p$, $p(Y)=p.v$ and the fact that  
\bes i_Y*(1)&=&\sqrt{-g}(v^0[123]-v^1[023]+v^2[013]-v^3[012])\\
&=& v_{\mu}*(dt^\mu),
\ees
we get
\bes
i_Y\Theta &=&\left(\frac{1}{2}[p_\mu(p.v)-v_\mu(p.p)]+\frac{1}{2}(p.v)p_\mu\right) *(dt^\mu)\\ 
&&-\frac{1}{2}m^2\phi^2v_\mu*(dt^\mu)\\
&=&\left.\left(p_\mu(p.v)-\frac{1}{2}\left[(p.p)
+m^2\phi^2)v_\mu \right]\right) *(dt^\mu)\right|_{0}\\
&=&
\la d\phi,Y^\flat\ra (*d\phi) -\frac{1}{2}\left[\la d\phi,d\phi\ra +m^2\phi^2\right](*Y^\flat)
\ees
which can also be written in the useful form
\be
i_Y\Theta &=&
\left[\phi_{,\mu}\phi_{,\nu} -\frac{1}{2}g_{\mu\nu}\left(g^{\alpha\beta}\phi_{,\alpha}\phi_{,\beta}
+m^2\phi^2\right)\right]v^\mu*(dt^\nu)\nonumber \\
&&
\ee

\section{Examples of conserved quantities}

As an illustration we calculate the conserved quantities for the Klein-Gordon field
in Minkowski background. In this case $L_Y\Theta=0$ (actually $\L_Y(*p\ww d\phi)=0$ and
$\L_Y H=0$ independently) for any of the ten Killing
vector fields $Y$ corresponding to Poincare transformations. 
For spacetime translations we have derived a formula in the last section.
Since we usually integrate on the 
spacelike surface $t^0=$ constant, it is enough to calculate the term
$*(dt^0)=-(123)$, which alone will give a non-zero contribution on $t=$  constant surface.
The following table gives the expected conserved 
quantities (energy and momentum densities) for time- and space-translations 
\vskip 5mm
\begin{center}
\begin{tabular}{ccl} $Y$ & $v^\mu$ & $-(123)$ part of $i_Y\Theta$\\ 
&&\\
$\ppp/\ppp t^0$ & $(1,0,0,0)$ & $(1/2)[(\phi_{,0})^2+(\nabla\phi)^2+m^2\phi^2]d^3t$\\
&&\\
$\ppp/\ppp t^1$ & $(0,1,0,0)$ & $[\phi_{,1}\phi_{,0}]d^3t$\\
\end{tabular}
\end{center}

\begin{appendix}
\section{Notation}
The spacetime is a Riemannian space with coordinates $t^\mu,\mu=0,1,2,3$.
Basis vectors in a tangent space are written $\ppp_\mu=\ppp/\ppp t^\mu$
The metric is given by the inner product 
$\la \ppp_\mu,\ppp_\nu\ra=g_{\mu\nu}$. The cotangent spaces have
basis elements $dt^\mu$ with $\la dt^\mu,dt^\nu\ra=g^{\mu\nu}$.
The metric has signature $(-1,1,1,1)$. The wedge product is defined so that
$\alpha\ww \beta=\alpha\tp\beta-\beta\tp\alpha$ for one-forms $\alpha$ and $\beta$.
The exterior derivative is defined so that for an $r$-form 
$\alpha=a_{\mu_1\dots\mu_r}dt^{\mu_1}\ww\dots\ww dt^{\mu_r}$ the derivative 
is the $(r+1)$-form
\bes d\alpha=a_{\mu_1\dots\mu_r,\nu}dt^\nu\ww dt^{\mu_1}\ww\dots\ww dt^{\mu_r}.\ees
The Hodge star is a linear operator that maps $r$-forms into $(4-r)$-forms
in our four-dimensional space.
The definition is 
\bes *(dt^{\mu_1}\ww\dots\ww dt^{\mu_r})&=&[(4-r)!]^{-1}\sqrt{-g}g^{\mu_1\nu_1}\dots\\
&&g^{\mu_r\nu_r}\varepsilon_{\nu_1\dots\nu_r\nu_{r+1}\dots\nu_4}dt^{\nu_{r+1}}\dots dt^{\nu_4}\ees
where $g$ denotes the determinant of $g_{\mu\nu}$ and $\varepsilon$ is the antisymmetric tensor
defined with $\varepsilon_{0123}=1$. The one-dimensional space of 0-forms has 
the unit vector equal to real number $1$. The one-dimensional space space of 
4-forms has the chosen orientation given by the unit vector $\varepsilon=n^0\ww n^1\ww n^2\ww n^3$
where $n^\mu$ are the orthonormal basis vectors. In ordinary basis
$\varepsilon=\sqrt{g}dt^0\ww dt^1\ww dt^2\ww dt^3$. The star operator acting on the zero
form equal to constant number $1$ is denoted by $*1=\varepsilon=\sqrt{g}dt^0\ww dt^1\ww dt^2\ww dt^3$.
We have the simple result that $dt^\mu\ww *dt^\nu=-*dt^\nu\ww dt^\mu=g^{\mu\nu}(*1)$

Note carefully that $*1$ is not the same as $*(1)$ where the shorthand notation $(\mu)$ is
used for $dt^\mu$. Similarly we use $(12)$ for $dt^1\ww dt^2$, $(013)$ for
$dt^0\ww dt^1\ww dt^3$ etc. 

The interior product $i(X)$ of a vector $X$ with an $r$-form $\alpha$ gives 
an $(r-1)$-form $i(X)\alpha$ defined by
\bes (i(X)\alpha)(Y_1,\dots,Y_{r-1})=\alpha(X,Y_1,\dots,Y_{r-1})\ees
When it is more convenient we will denote the interior product operator by $i_X$
in place of $i(X)$.

Two successive applications of interior products on a form will be denoted by
\bes i(X,Y)\alpha\equiv [i(X)\circ i(Y)]\alpha = i(X)[i(Y)\alpha] \ees
Note that $i(X,Y)=-i(Y,X)$. Similarly successive applications $i(XY\dots Z)$ of many such
interior products can be defined. If $\alpha$ is an $r$-form then
\bes i(X)(\alpha\ww\beta)=[i(X)\alpha]\ww\beta+(-1)^r\alpha\ww i(X)\beta\ees
In order to abbreviate expressions we use $i(12)$ for $i(X_1X_2)=i(X_1)\circ i(X_2)$ etc. when there is no
confusion.

\section{Solution submanifold of $\Omega$}
We give the calculation of $i(X_3X_2X_1X_0)\Omega$ for $H=(*p)\ww p/2+m^2\phi^2*(1)/2$
in Minkowski space for illustration. We take
the independent tangent vectors to the section 
\bes \sigma : t\to (t^\mu, \phi=F(t),p_\nu=G_\nu(t)) \ees
the push-forwards
\bes X_\mu\equiv \sigma_*(\ppp_\mu)=\ppp_\mu+F_{,\mu}\ppp_\phi+G_{\nu,\mu}\ppp_{p_\nu}\ees 
The calculation involves the following expressions (we use abbreviations of Appendix A) :
\bes *1 &=& (0123)\\ 
*dt^\mu &=& [-(123),-(023),+(013),-(012)]\ees
\bes
d(p_\mu *dt^\mu)&=& dp_\mu *dt^\mu=-dp_0(123)-dp_1(023)+dp_2(013)-dp_3(012)\\
i(0)(0123)&=& (123),\\ i(1)(0123) &=& -(023),\\ i(2)(0123) &=& (013),\\ i(3)(0123)&=&-(012)\\
&&\\
i(0)(d*p)& = & dp_1(23)-dp_2(13)+dp_3(12)-G_{0,0}(123)-G_{1,0}(023)\\&&+G_{2,0}(013)-G_{3,0}(012)\\
i(1)(d*p)& = & dp_0(23)+dp_2(03)-dp_3(02)-G_{0,1}(123)-G_{1,1}(023)\\&&+G_{2,1}(013)-G_{3,1}(012)\\
i(2)(d*p)& = & -dp_0(23)-dp_1(03)+dp_3(01)-G_{0,2}(123)-G_{1,2}(023)\\&&+G_{2,2}(013)-G_{3,2}(012)\\
i(0)(d*p)& = & dp_0(12)+dp_1(02)-dp_2(01)-G_{0,3}(123)-G_{1,3}(023)\\&&+G_{2,3}(013)-G_{3,3}(012)\\
i(10)(d*p)&=&dp_2(3)-dp_3(2)-G_{0,0}(23)-G_{2,0}(03)+G_{3,0}(02)\\&&+G_{1,1}(23)-G_{2,1}(13)+G_{3,1}(12)\\
i(20)(d*p)&=&-dp_1(3)+dp_3(1)+G_{0,0}(13)+G_{1,0}(03)-G_{3,0}(01)\\&&+G_{1,2}(23)-G_{2,2}(13)+G_{3,2}(12)\\
i(30)(d*p)&=&dp_1(2)-dp_2(1)-G_{0,0}(12)-G_{1,0}(02)+G_{2,0}(01)\\&&+G_{1,3}(23)-G_{2,3}(13)+G_{3,3}(12)\\
i(21)(d*p)&=&-dp_0(3)-dp_3(0)+G_{0,1}(13)+G_{1,1}(03)-G_{3,1}(01)\\&&+G_{0,2}(23)+G_{2,2}(03)-G_{3,2}(02)\\
i(31)(d*p)&=&dp_0(2)+dp_2(0)-G_{0,1}(12)-G_{1,1}(02)+G_{2,1}(01)\\&&+G_{0,3}(23)+G_{2,3}(03)-G_{3,3}(02)\\
i(32)(d*p)&=&-dp_0(1)-dp_1(0)-G_{0,2}(12)-G_{1,2}(02)+G_{2,2}(01)\\&&-G_{0,3}(13)-G_{1,3}(03)+G_{3,3}(01)\\
i(321)(d*p)&=&dp_0-G_{0,1}(1)-G_{1,1}(0)-G_{0,2}(2)-G_{2,2}(0)\\&&-G_{0,3}(3)-G_{3,3}(0)\\
i(320)(d*p)&=&dp_1-G_{0,0}(1)-G_{1,0}(0)-G_{1,2}(2)+G_{2,2}(1)\\&&-G_{1,3}(3)+G_{3,3}(1)\\
i(310)(d*p)&=&-dp_2+G_{0,0}(2)+G_{2,1}(0)-G_{1,1}(2)+G_{2,1}(1)\\&&+G_{2,3}(3)-G_{3,3}(2)\\
i(210)(d*p)&=&dp_3-G_{0,0}(3)-G_{3,0}(0)+G_{1,1}(3)-G_{3,1}(1)\\&&+G_{2,2}(3)-G_{3,2}(2)\\
i(3210)(d*p)&=&-G_{0,0}+G_{1,1}+G_{2,2}+G_{3,3}\\
\ees
If $A$ is a 4-form and $B$ a 1-form then
\bes i(3210)[A\ww B]&=&[i(3210)A]B+[i(321)A]i(0)B-[i(320)A]i(1)B\\&&+[i(310)A]i(2)B-[i(210)A]i(3)B \ees
For $A=d*p-m^2\phi *1$ and $B=p-d\phi$ the expression for $i(3210)[A\ww B]=i(3210)\Omega$ is
a 1-form in the extended phase space which should be equated to zero. The coefficients of
$dp_\mu$ equated to zero give $G_\mu-F_{,\mu}=0$ and the coefficient of $d\phi$ 
gives  $-G_{0,0}+G_{1,1}+G_{2,2}+G_{3,3}=0$. These imply the Klein-Gordon equation for
$F$. 
\end{appendix}

\end{document}